\documentclass[dvips]{acta}
\usepackage{supertabular,lscape,epsfig}
\usepackage{amssymb}
\usepackage{amsmath}
\usepackage{wasysym}
\DeclareSymbolFont{ppa}{OT1}{ppl}{m}{it}
\DeclareMathSymbol{\vv}{\mathalpha}{ppa}{'166}

\newfont{\hb}{rphvb at 10pt}
\newfont{\hbo}{rphvbo at 10pt}
\newfont{\bitt}{rptmbi at 12pt}
\newfont{\bits}{rptmbi at 11pt}

\SetPages{0}{0}

\SetVol{0}{0}

\begin{document}

\newcommand{\TabCapp}[2]{\begin{center}\parbox[t]{#1}{\centerline{
  \small {\spaceskip 2pt plus 1pt minus 1pt T a b l e}
  \refstepcounter{table}\thetable}
  \vskip2mm
  \centerline{\footnotesize #2}}
  \vskip3mm
\end{center}}

\newcommand{\TTabCap}[3]{\begin{center}\parbox[t]{#1}{\centerline{
  \small {\spaceskip 2pt plus 1pt minus 1pt T a b l e}
  \refstepcounter{table}\thetable}
  \vskip2mm
  \centerline{\footnotesize #2}
  \centerline{\footnotesize #3}}
  \vskip1mm
\end{center}}

\newcommand{\MakeTableSepp}[4]{\begin{table}[p]\TabCapp{#2}{#3}
  \begin{center} \TableFont \begin{tabular}{#1} #4 
  \end{tabular}\end{center}\end{table}}

\newcommand{\MakeTableee}[4]{\begin{table}[htb]\TabCapp{#2}{#3}
  \begin{center} \TableFont \begin{tabular}{#1} #4
  \end{tabular}\end{center}\end{table}}

\newcommand{\MakeTablee}[5]{\begin{table}[htb]\TTabCap{#2}{#3}{#4}
  \begin{center} \TableFont \begin{tabular}{#1} #5 
  \end{tabular}\end{center}\end{table}}

\newfont{\bb}{ptmbi8t at 12pt}
\newfont{\bbb}{cmbxti10}
\newfont{\bbbb}{cmbxti10 at 9pt}
\newcommand{\uprule}{\rule{0pt}{2.5ex}}
\newcommand{\douprule}{\rule[-2ex]{0pt}{4.5ex}}
\newcommand{\dorule}{\rule[-2ex]{0pt}{2ex}}

\renewcommand{\labelitemi}{$\diamond$}

\begin{Titlepage}
\Title{Open clusters in 2MASS photometry \\ II. Mass Function and Mass Segregation}

\Author{{\L}. ~B~u~k~o~w~i~e~c~k~i$^1$, ~G. ~M~a~c~i~e~j~e~w~s~k~i$^1$, ~P. ~K~o~n~o~r~s~k~i$^2$ ~and ~A. ~N~i~e~d~z~i~e~l~s~k~i$^1$}
{$^1$Toru\'n Centre for Astronomy, Nicolaus Copernicus University, Gagarina 11, PL-87-100 Toru\'n, Poland\\
$^2$Warsaw University Observatory, Al. Ujazdowskie 4, 00-478, Warsaw, Poland\\~\\e-mail: bukowiecki@astri.umk.pl}

\Received{Month Day, Year}
\end{Titlepage}

\Abstract{This is a continuation of our study of open clusters based on the 2--Micron All Sky Survey photometry. Here we present the results of the mass function analysis for 599 known open clusters in the Milky Way. The main goal of this project is a study of the dynamical state of open clusters, the mass segregation effect and an estimate of the total mass and the number of cluster members. We noticed that the cluster size (core and overall radii) decreases along dynamical evolution of clusters. The cluster cores evolve faster than the halo regions and contain proportionally less low-mass stars from the beginning of the cluster dynamical evolution. We also noticed, that the star density decreases for the larger clusters. Finally, we found an empirical relation describing the exponential decrease of the mass function slope with the dynamical evolution of clusters.}{open clusters and associations: general -- infrared: galaxies -- astronomical databases: 2MASS -- stars: mass function}


\section{Introduction}
The distribution of the stellar masses that stars were formed with, can be described as an empirical function -- the Initial Mass Function (IMF, Salpeter 1955, Miller \& Scalo 1979). According to the basic considerations, the IMF is expected to depend on the star forming conditions (Larson 1998). However, the studies presented more recently indicate that the mass distribution is relatively invariant from one open cluster to another and has a universal character (Phelps \& Janes 1993; Massey \etal 1995; Durgapal \& Pandey 2001; Kroupa 2001, 2002; Pandey \etal 2005, 2007; Sharma \etal 2008). Similar to our research, Bica \& Bonatto (2005) and Bonatto \& Bica (2005) used the $JHK_{S}$  photometry data to study the mass function (MF) of 5 and 11 open clusters, respectively, and pointed out that the MF slope decreases along the dynamical evolution of the investigated clusters. Maciejewski \& Niedzielski (2007) used homogeneous $BV$ data for 42 open clusters to analyze the MF and observed a similar trend.

This paper is a continuation of our study of known open clusters (Bukowiecki \etal 2011, hereafter Paper I). From our sample of 849 open clusters, we were able to analyze the MF for 599 clusters. We used the near--infrared $JHK_{S}$ photometric data to determine the initial mass function $(\chi_{0})$, the dynamical evolution, the mass of the cluster and the number of member stars. In Section 2, the method and the data analysis is presented. In Section 3, our results are compared to the results found in the literature. Section 4 contains discussion of the relations between individual parameters. Final conclusions are summarized in Section 5.


\section{Data analysis}

\subsection{Data Source and Cluster Selection}
Our research is based on the $JHK_{S}$ photometric data extracted from the 2MASS\footnote{http://www.ipac.caltech.edu/2mass/releases/allsky/} \emph{Point Source Catalog} (Skrutskie \etal 2006). The method is described in Paper I. For a sample of 849 open clusters we determined new coordinates of the centers and the angular sizes. Moreover, age, reddening, distance, and linear sizes were also derived for 754 of them (Table 1 and 2 in Paper I).

We studied only the MF for a part of the whole sample because of two reasons. First, we had to reject clusters without a complete set of parameters. Second, after a series of tests, we decided to study clusters for which we observe more than two magnitude of the main sequence, between the turnoff point and a level of 15.8 mag in the $J$ band -- a value of the $99.9\%$ Point Source Catalogue Completeness Limit\footnote{Following the Level 1 Requirement, according to Explanatory Supplement to the 2MASS All Sky Data Release and Extended Mission Products (http://www.ipac.caltech.edu/2mass/releases/allsky/doc)}. That allowed us to reach a reasonable precision of estimated parameters. After applying these criteria, we included 599 open clusters in the further MF analysis.

\subsection{Mass Functions and the Dynamical--Evolution Parameter}
The first step to derive the MF $\phi(m)$ was to build the cluster's luminosity function (LF). After a series of tests we used 0.5 mag bins, for the core, the corona (halo), and the overall regions separately. Another LF was built for an offset field starting at $r = r_{lim}  + 1$ arcmin (where $r_{lim}$ is the limiting size of the open cluster, and generally "$_{lim}$" is related to the overall cluster). The LF of the offset field was subtracted, bin by bin, from the cluster LF, taking the area proportion into account. The resulting LF was converted into the MF using a respective isochrone. This way we derived the MF approximated by a standard relation of the form
\begin{equation}
\centering
\text{log}\: \phi (m) = \text{log}(dN/dm) = -(1+\chi)\:\text{log}\:m + b_{0}\:,
\end{equation}
\begin{figure}[b]
\begin{center}
\includegraphics[width=1.0\textwidth]{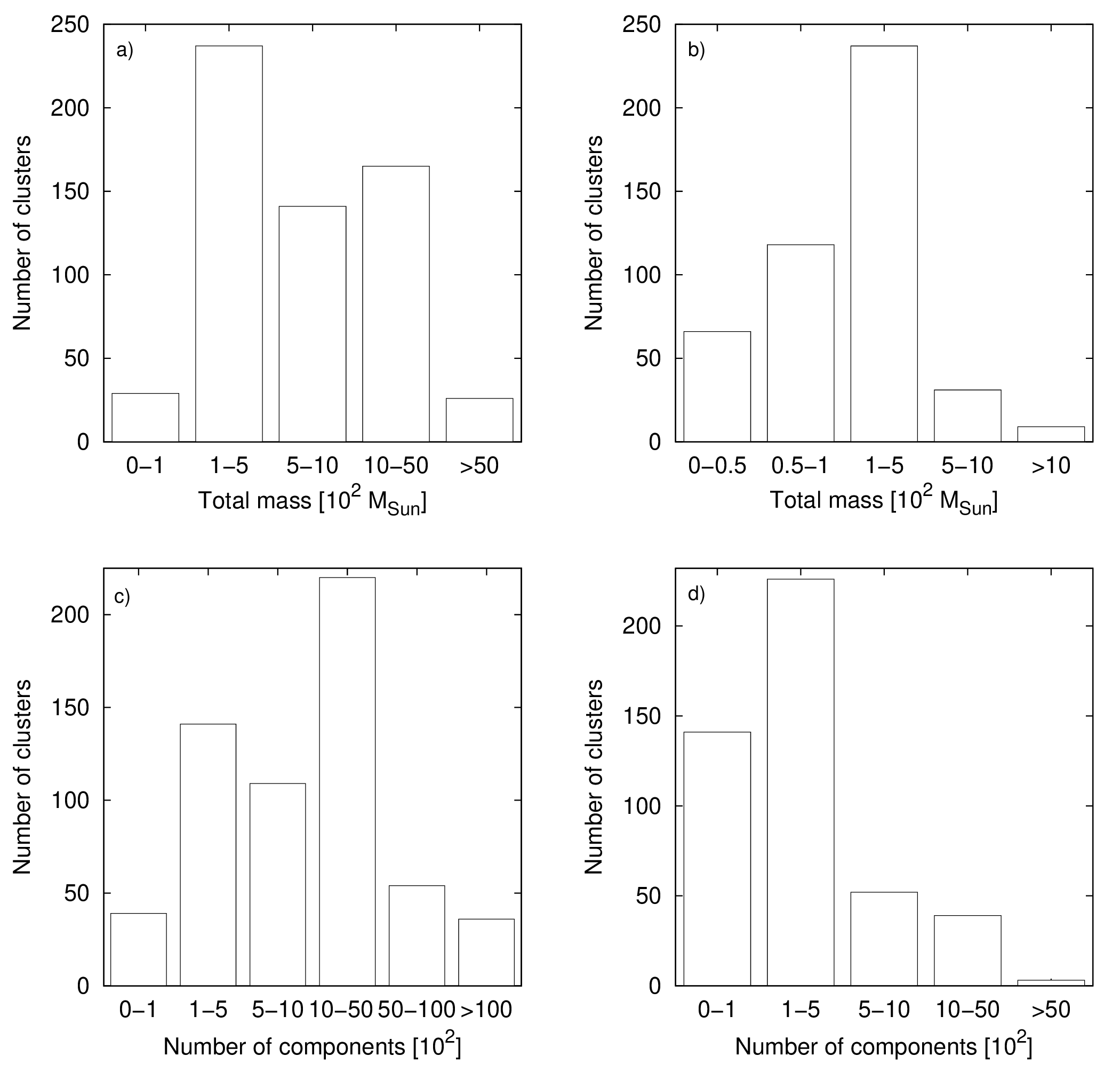}
\end{center}
\FigCap{$a)$ Number of clusters in the distribution of the total mass of clusters. $b)$ The same as $panel\:\: a$ but for core radii. $c)$ Number of clusters in the distribution of the total number of components. $d)$ The same as $panel\:\: c$ but for core radii.}
\end{figure}
where $m$, $N$, $\chi$, and $b_{0}$ are the stellar mass, the number of stars, the mass function slope, and a constant, respectively. Using basic astrophysical parameters taken from Paper I we estimated the additional cluster parameters: the total mass $M_{total}$, total number of stars $N_{total}$, the core mass $M_{core}$ and the number of stars in the core $N_{core}$ (Fig.~1). Based on our calculations we concluded that 91\% of the clusters in the sample have a total mass in the range of 100 -- 5000 $M_{\odot}$ and the mass of the core for 84\% clusters is in the range of 50 -- 1000 $M_{\odot}$, while 78\% of studied clusters contain between 100 and 5000 stars. Cores are naturally smaller, 80\% of the clusters have less than 500 stars within them. These values were calculated by extrapolating the MF from the star-mass limit of $0.08 M_{\odot}$ to the turnoff using the methods described in Bica \& Bonatto (2005), if $\chi$ was lower than the universal initial mass function, $\chi_{IMF} =1.3 \pm 0.3$ (Kroupa 2001). In the other case the MF was extrapolated with  $\chi = 0.3$ from $0.08 M_{\odot}$ to the mass of $0.5 M_{\odot}$ and for greater stellar masses with the derived $\chi$ to the turnoff. This procedure was applied to analysis of the overall MF for the whole sample of 599 open clusters. In addition, for 461 clusters, with $r_{core} > 0.6\:'$, we studied the MF for the core and halo regions separately. The error bars for $\phi(m)$ were calculated assuming the Poisson statistics. The obtained parameters are listed in Table~1.

Section 2 of Paper I describes the procedure of estimating size, age and distance in details.

To describe the dynamic aspect of the studied open clusters, we used the dynamical-evolution parameter $\tau$, defined as
\begin{equation}
\centering
\tau = \frac{age}{t_{relax}} \:,
\end{equation}
where the relaxation time $t_{relax}$ was calculated as
\begin{equation}
\centering
t_{relax} = \frac{N}{8\:ln\:N}\:t_{cross} \:.
\end{equation}
Here $t_{cross} = R/\sigma_{\nu}$ is the crossing time, $N$ is the number of stars in studied radius $R$ and $\sigma_{\nu}$ is the velocity dispersion (Binney \& Tremaine 1987) with a typical value of 3 km s$^{-1}$ (Binney \& Merrifield 1998). Our calculations were done for the entire region of the open cluster and separately for the core and halo areas. To describe differences between the MF slope of the core and the coronal regions we used $\Delta \chi = \chi_{halo} - \chi_{core}$, which can be treated as the mass segregation indicator.

\subsection{Structural Parameters}

\begin{figure}
\begin{center}
\includegraphics[width=0.9\textwidth]{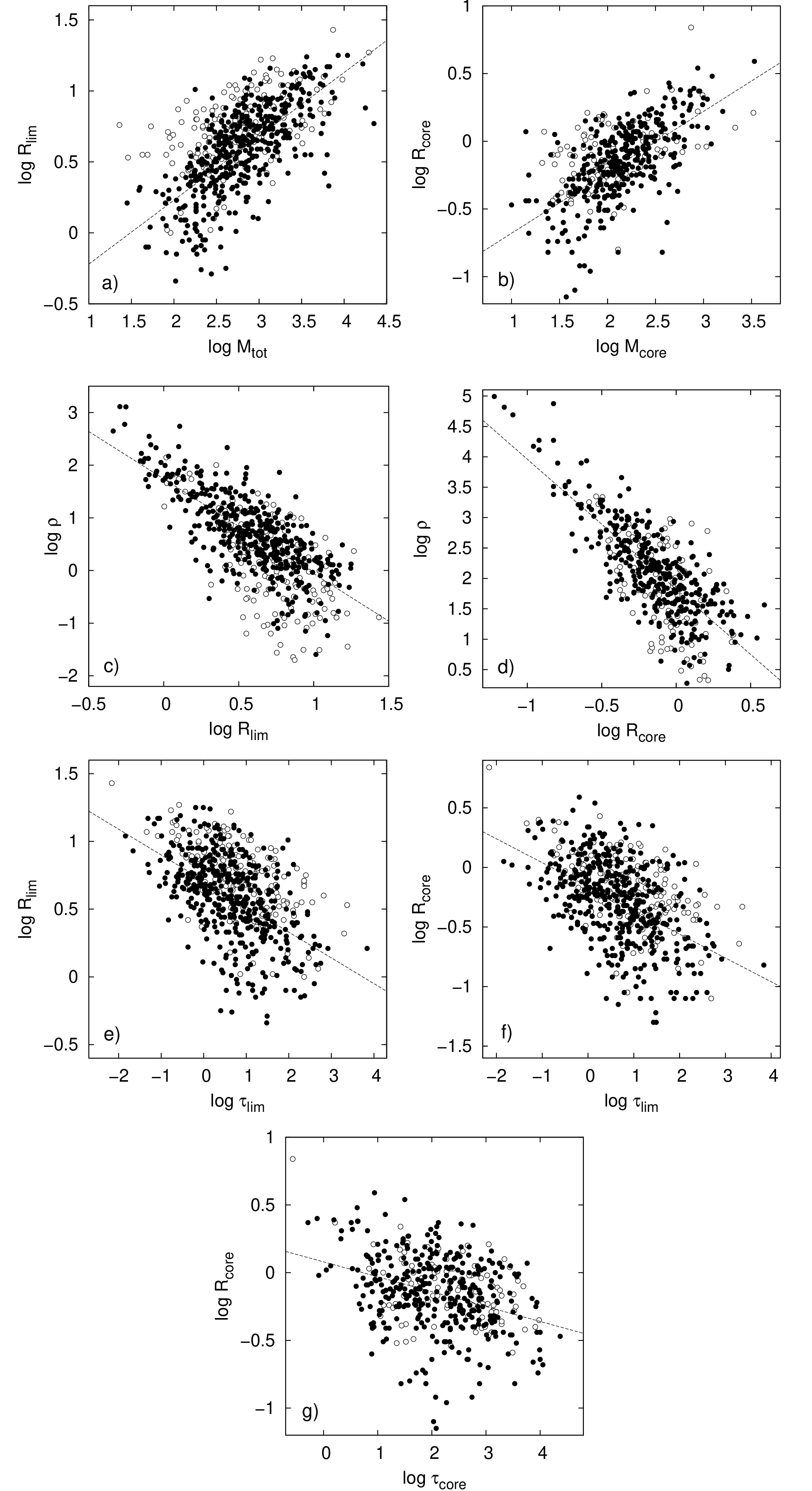}
\end{center}
\FigCap{Relations between cluster structural parameters. Full circles represent open clusters used for fitting a linear relation and empty circles are determinations with errors greater than 100\% of a derived value, which have not been taken into account. See Sect. 2.3 for details.}
\end{figure}

In this section we present the relations of the open cluster's size with the cluster's mass and the dynamic evolution parameter. From the total sample of studied clusters, we analyzed only those for which the relative uncertainty of the total mass or the number of potential members was smaller than 100\% of the derived value. The determinations with greater errors, presented as the open points in all Figures, were omitted in further analysis. Analyzing the total mass of 436 clusters (panel $a$ of Fig.~2), we obtained a linear relation:
\begin{equation}
\centering
\text{log}\: R_{lim} = (0.45 \pm 0.03)\:\text{log}\:M_{total} - (0.67 \pm 0.07) \:,
\end{equation}
with the correlation coefficient of 0.70. As one could expect, the core radius is also related to the core mass. In this case for 351 clusters (panel $b$ of Fig.~2), we fitted a linear trend that resulted in an equation:
\begin{equation}
\centering
\text{log}\: R_{core} = (0.45 \pm 0.03)\:\text{log}\:M_{core} - (1.13 \pm 0.07) \:,
\end{equation}
the correlation coefficient was found to be 0.64. These both trends indicate that the clusters with small diameters have small mass, and vice versa. A similar conclusion was reached by Maciejewski \& Niedzielski (2007).

We plotted the star density $\rho$ (stars/parsec$^3$) in panels $c$ and $d$ of Fig.~2 as a function of the open cluster's diameter. For the clusters with determined limiting (431 objects) and core (350 objects) parameters (the mass and diameter), we found linear relations:
\begin{equation}
\centering
\text{log}\: \rho_{lim} = (-1.80 \pm 0.08)\:\text{log}\:R_{lim} + (1.74 \pm 0.06) \:,
\end{equation}
\begin{equation}
\centering
\text{log}\: \rho_{core} = (-2.14 \pm 0.09)\:\text{log}\:R_{core} + (1.82 \pm 0.03) \:,
\end{equation}
with the correlation coefficients equal to 0.75 and 0.80, respectively. This indicates that the star density of open clusters decreases with their diameters. For the smaller clusters $\rho_{lim}$ is much greater than for the larger ones. It is worth noting that for the core regions this trend is more significant. This finding suggests that the migration of stars in the central regions is stronger. No relation between $\rho$ and the number of cluster members, mass or age was found in the investigated sample.

The dynamical evolution is described by the $\tau$ parameter. We used 431 (panels $e$ and $f$ of Fig.~2) and 350 clusters (panel $g$ of Fig.~2) to investigate how the open cluster's core and limiting radii depend on $\tau$. We derived the following relations:
\begin{equation}
\centering
\text{log}\: R_{lim} = (-0.19 \pm 0.02)\:\text{log}\:\tau_{lim} + (0.71 \pm 0.02) \:,
\end{equation}
\begin{equation}
\centering
\text{log}\: R_{core} = (-2.20 \pm 0.02)\:\text{log}\:\tau_{lim} + (-0.16 \pm 0.02) \:,
\end{equation}
\begin{equation}
\centering
\text{log}\: R_{core} = (-0.11 \pm 0.02)\:\text{log}\:\tau_{core} + (0.08 \pm 0.04) \:,
\end{equation}
resulting in the weak relation with correlation coefficients of 0.53, 0.50 and 0.35, respectively. These results demonstrate that open clusters tend to decrease in size during their dynamic evolution, both the overall and the core radii. Similar results were reported by Maciejewski \& Niedzielski (2007) and Nilakshi \etal (2002), who noted a decrease in the size for the older and the dynamically more evolved systems.

No relation between the concentration parameter, defined as $c =$ log $(R_{lim}/R_{core})$ (Peterson and King 1975) and $\tau$, age or mass segregation was found in the studied sample.
\begin{figure}
\begin{center}
\includegraphics[width=1.0\textwidth]{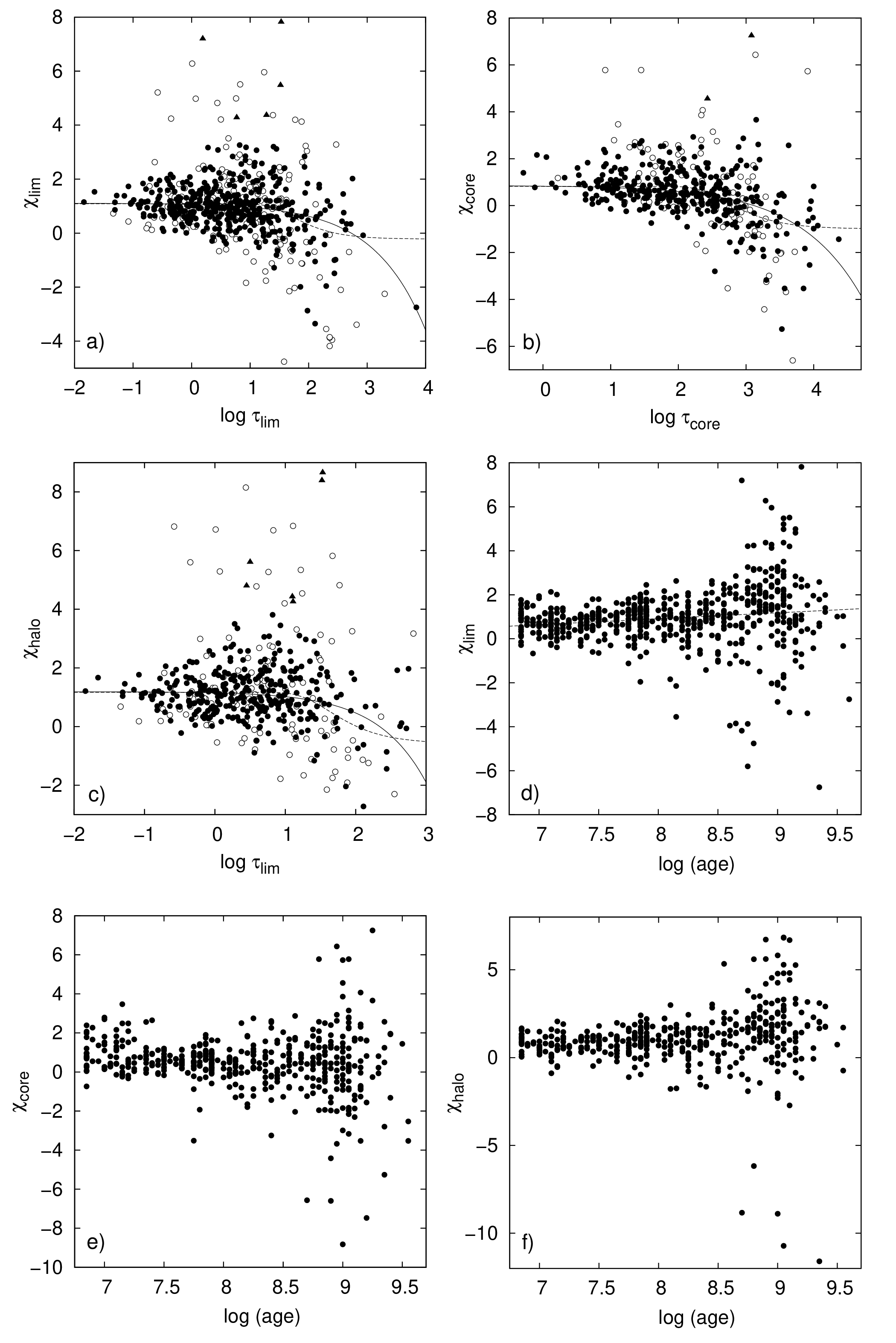}
\end{center}
\FigCap{Relations between the MF slopes and other clusters parameters. Full circles represent open clusters used for fitting relations $\chi(\tau) = \chi_{0}-\chi_{1} \: \text{exp}\: (\text{log} \: \tau/\tau_{0})$ using the solid line and $\chi(\tau) = \chi_{0}-\chi_{1} \: \text{exp}\: (- \tau_{0}/\tau)$ with the dotted line ($panels\:\: a,b$ and $c$) or linear relation ($panel\:\: d$). While triangles are outlying points and empty circles are measurements with errors greater than 100\% of a derived value, both (empty circles and triangles) have not been taken into account during fitting functions.}
\end{figure}

\subsection{Mass Function Slopes}
The MF slope is characterized by the $\chi$ parameter and describes the mass distribution in an open cluster. It can be interpreted as a measure of the estimated quantities of the low-mass members in comparison to the more massive stars. The small $\chi$ guarantees the deficit of the low-mass stars in the cluster. And, vice versa, the higher $\chi$ illustrate a surplus of low-mass components. We also inspected relations between the MF slope and the dynamical-evolution parameter $\tau$. Bonatto \& Bica (2005) obtained an empirical relation between $\chi$ and $\tau$ in the form of $\chi(\tau) = \chi_{0} - \chi_{1} \:\text{exp}(-\tau_{0}/\tau)$ (hereafter the B\&B relation; $\chi_{0}$ is the initial mass function of studied clusters) which suggests the MF slope to decrease exponentially with $\tau$. Our results do not confirm this relation. We noticed that the MF slope tends to decrease for high values of $\tau$. Thus our investigation led to another empirical relation describing the exponential decrease more precisely: $\chi(\tau) = \chi_{0} - \chi_{1} \:\text{exp}\:(\text{log} \: \tau / \tau_{0})$.

For $\chi_{lim}$ and $\tau_{lim}$ (panel $a$ of Fig.~3), using the least-square fit to 426 clusters and after removing five outlying points, we obtained the following relation:
\begin{equation}
\centering
\chi_{\:lim}(\tau) = (1.11 \pm 0.05) - (0.05 \pm 0.02)\:\text{exp}(\text{log} \: \tau_{lim} / 1.88 \pm 0.10) \:,
\end{equation}
with the correlation coefficients of 0.41 and reduced $\chi^2 = 0.89$, while for the B\&B relation fitted to our data we obtained:
\begin{equation}
\centering
\chi_{\:lim}(\tau) = (1.09 \pm 0.05) - (1.31 \pm 0.24)\:\text{exp}(-52.5 \pm 17.6 / \tau_{lim}) \:,
\end{equation}
with the correlation coefficients of 0.39 and reduced $\chi^2 = 0.92$.

For $\chi_{core}$ and $\tau_{core}$ (panel $b$ of Fig.~3), we used 348 clusters and after removing two outlying points, we obtained:
\begin{equation}
\centering
\chi_{\:core}(\tau) = (0.86 \pm 0.06) - (0.03 \pm 0.02)\:\text{exp}(\text{log} \: \tau_{core} / 0.93 \pm 0.12) \:,
\end{equation}
with the correlation coefficient equal to 0.45 and reduced $\chi^2 = 0.88$, while for the B\&B relation, fitted to our data:
\begin{equation}
\centering
\chi_{\:core}(\tau) = (0.82 \pm 0.05) - (1.81 \pm 0.22)\:\text{exp}(-520 \pm 116 / \tau_{core}) \:,
\end{equation}
with the correlation coefficient equal to 0.42 and reduced $\chi^2 = 0.93$.

For $\chi_{halo}$ and $\tau_{lim}$ (panel $c$ of Fig.~3), using 328 clusters and after removing six outlying points, the least-square fitted relations are:
\begin{equation}
\centering
\chi_{\:halo}(\tau) = (1.19 \pm 0.07) - (0.04 \pm 0.03)\:\text{exp}(\text{log} \: \tau_{lim} / 0.69 \pm 0.14) \:,
\end{equation}
with the correlation coefficient equal to 0.36 and reduced $\chi^2 = 0.97$, while for the B\&B relation:
\begin{equation}
\centering
\chi_{\:halo}(\tau) = (1.17 \pm 0.07) - (1.75 \pm 0.36)\:\text{exp}(-42.0 \pm 14.8 / \tau_{lim}) \:,
\end{equation}
with the correlation coefficient equal to 0.34 and reduced $\chi^2 = 1.02$. In all these cases our new relation fit better than the B\&B one.

It is worth mentioning that the obtained value of $\chi_{0}$ for both overall and halo cluster regions are similar to $\chi_{IMF} = 1.3 \pm 0.3$ (Kroupa 2001) while the $\chi_{0}$ for the core is noticeably smaller. The same finding was presented in Maciejewski \& Niedzielski (2007) and Bonatto \& Bica (2005). This suggests that the core regions contain proportionally less low-mass stars than the halo areas what may be interpreted as a mass segregation effect. For the core region both relations correlate better, indicating that halo regions are treated in some processes that make it difficult to study (separating, mixing or errors in estimate of the cluster size). Moreover, the decrease of the MS slope indicates that the evaporation of the low-mass members occurs in the entire cluster. Also the more massive stars might migrate to the core from the halo region.

We noticed weak (the correlation coefficient equal to 0.15) connection between the MF slope for overall (panel $d$ of Fig.~3):
\begin{equation}
\centering
\chi_{\:lim} = (0.27 \pm 0.08)\:\text{log}(age) + (-1.25 \pm 0.62)\:,
\end{equation}
No relations between the MF slope for core (panel $e$ of Fig.~3) or coronal (panel $f$ of Fig.~3) areas and cluster age were detected in the investigated sample. However, in the open clusters older than about 500 Myr we noticed that $\chi$ for all mentioned regions becomes either very high or low as compared to the mean value.
\begin{figure}
\begin{center}
\includegraphics[width=0.9\textwidth]{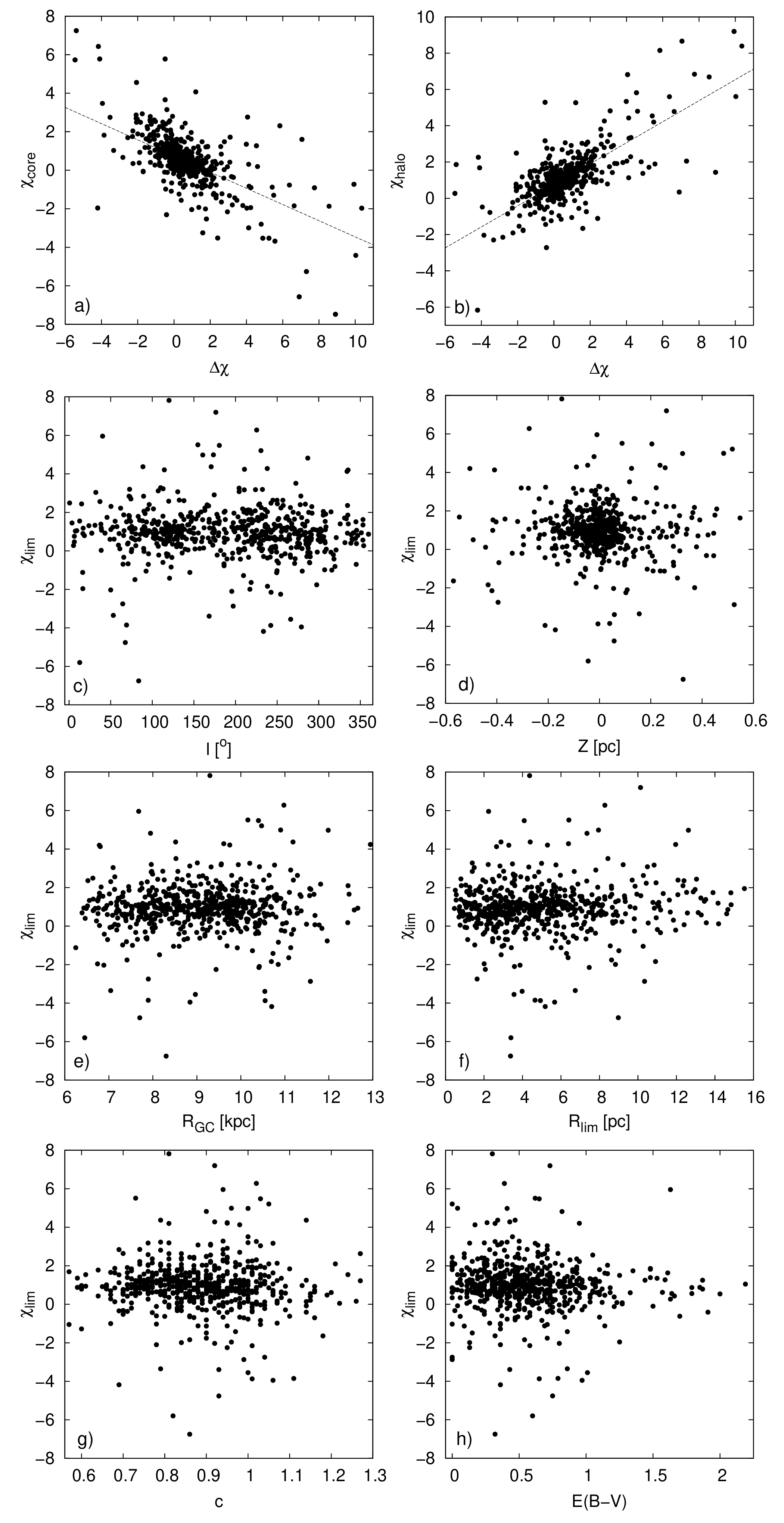}
\end{center}
\FigCap{All panels show relations of the MF slopes with a spatial and structural parameters derived in Paper I. See Sect. 2.4 for details.}
\end{figure}

As one can see in panel $a$ of Fig.~4, $\chi_{core}$ is correlated with the mass segregation parameter $\Delta \chi$. For 461 open cluster, we obtained the relation:
\begin{equation}
\centering
\chi_{\:core} = (-0.42 \pm 0.03)\:\Delta \chi + (0.74 \pm 0.06)\:,
\end{equation}
with the correlation coefficient equal to 0.64. As shown in panel $b$ of Fig.~4, $\chi_{halo}$ depends on $\Delta \chi$, as well. While considering this same number of clusters in the previous case, we derived a linear relation:
\begin{equation}
\centering
\chi_{\:halo} = (0.58 \pm 0.03)\:\Delta \chi + (0.74 \pm 0.06)\:,
\end{equation}
with the correlation coefficient of 0.76.

\begin{figure}[t]
\begin{center}
\includegraphics[width=0.9\textwidth]{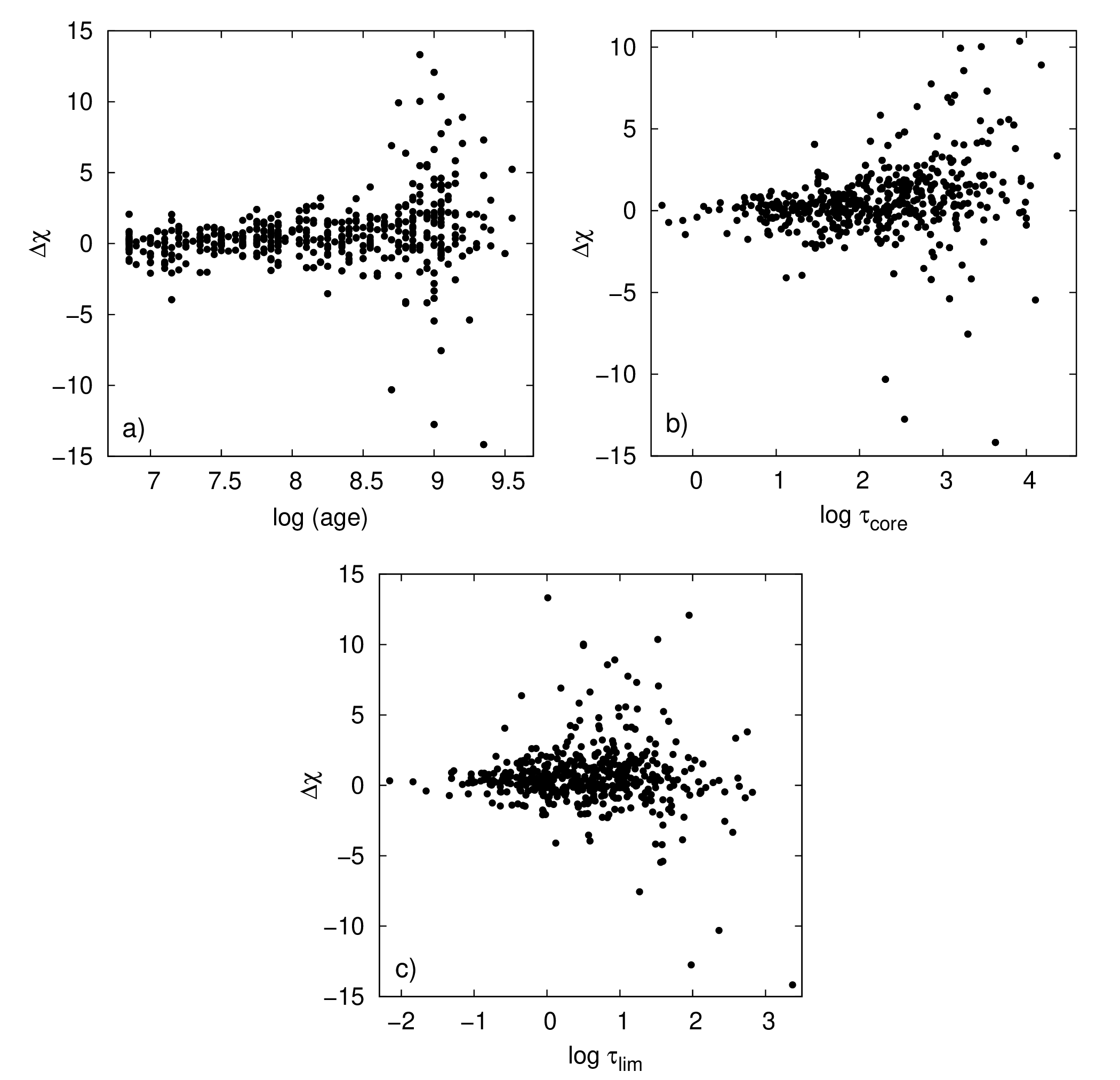}
\end{center}
\FigCap{Evolution of the mass segregation measure in time.}
\end{figure}
\begin{figure}[t]
\begin{center}
\includegraphics[width=1.0\textwidth]{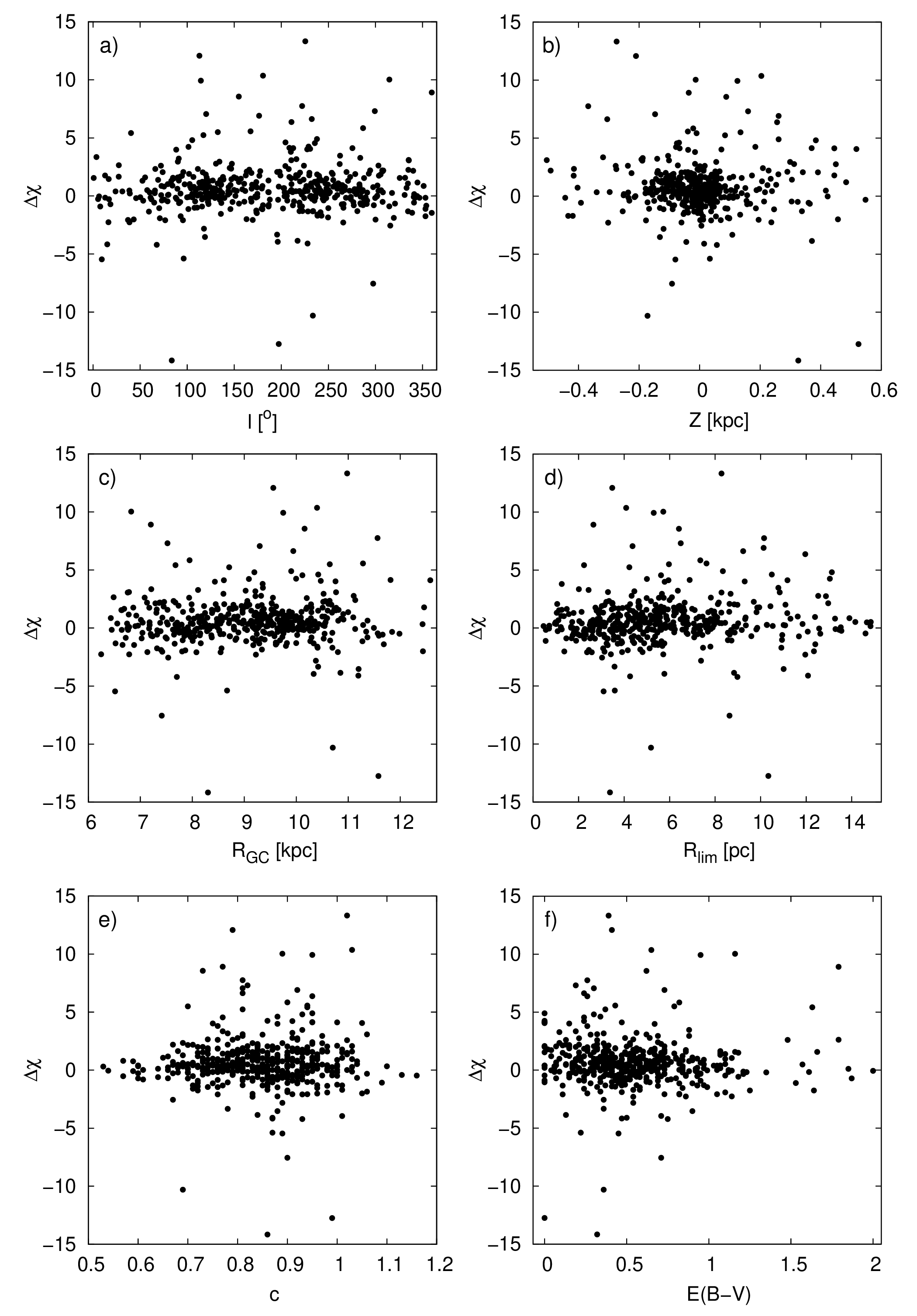}
\end{center}
\FigCap{All panels show the mass segregation versus location in the Galaxy and structural parameters derived in Paper I. See Sect. 2.5 for details.}
\end{figure}

In panels $c,d$ and $e$ of Fig.~4 we show that the MF slope is not related with the location in our Galaxy such as the Galactic Longitude ($l$), the distance from the Galactic plane ($Z$) or the Galactic center ($R_{GC}$). Panels $f$ and $g$ of Fig.~4 show no relations between $\chi_{lim}$ and open clusters' structural parameters such as the limiting radii ($R_{lim}$) or the concentration parameter ($c$). Moreover, (panel $h$ of Fig.~4) we did not find any relation between the reddening and the MF slope.

\subsection{Mass Segregation}
The dispersion of the MS slope versus the age, similar to that shown in panels $d,e$ and $f$ of Fig.~3, can be observed in the comparison presented in panel $a$ of Fig.~5. Moreover the mass segregation tends to increase with age, which suggests the existence of the initial mass segregation within the protostellar gas cloud, as noticed by Maciejewski \& Niedzielski (2007) and Sharma \etal (2008).

The following panel $b$ of Fig.~5 indicates that the mass segregation $\Delta \chi$ increases with the dynamic evolution of the core. This in turn suggests that the core regions evolve faster than the halo regions so the effect of the evaporation of the low-mass members is stronger there. Another process which might be responsible for the mentioned increases, is the migration of the more massive components from the halo regions to the core. This suggests that the dynamical-evolution parameter $\tau$ is a better indicator describing the evolution status of open clusters than the cluster age. According to equation (2), a small open cluster with small number of members can be more evolved than a huge cluster with many components. In panel $c$ of Fig.~5 we plotted $\Delta \chi$ as a function of the dynamical-evolution parameter $\tau$. No relation can be seen, however only for clusters older than their relaxation time, i.e. with log$ \:\tau > 0$ we observe a strong mass segregation. A similar results was obtained by Maciejewski \& Niedzielski (2007).

In panels $a-f$ of Fig.~6 we present the mass segregation plotted against the location in the Galaxy and the structural parameters of open clusters derived in Paper I. No statistically relations between these parameters were found.

\section{Comparison with the Published Data}
To test the reliability of our determination, we compared our results with the literature data. We collected the literature data from various papers dedicated to individual clusters, when available. In panels $a$ and $b$ of Fig.~7 (for 63 and 42 open clusters, respectively) presents our MF slope plotted against the published values. After rejecting one and six outlying points, respectively, we obtained the linear relations:
\begin{equation}
\centering
\chi^{\:our}_{\:lim} = (0.96 \pm 0.05)\: \chi^{\:lit}_{\:lim}\:
\end{equation}
\begin{equation}
\centering
\chi^{\:our}_{\:core} = (0.87 \pm 0.07)\: \chi^{\:lit}_{\:core}\:
\end{equation}
\begin{figure}[t]
\begin{center}
\includegraphics[width=0.9\textwidth]{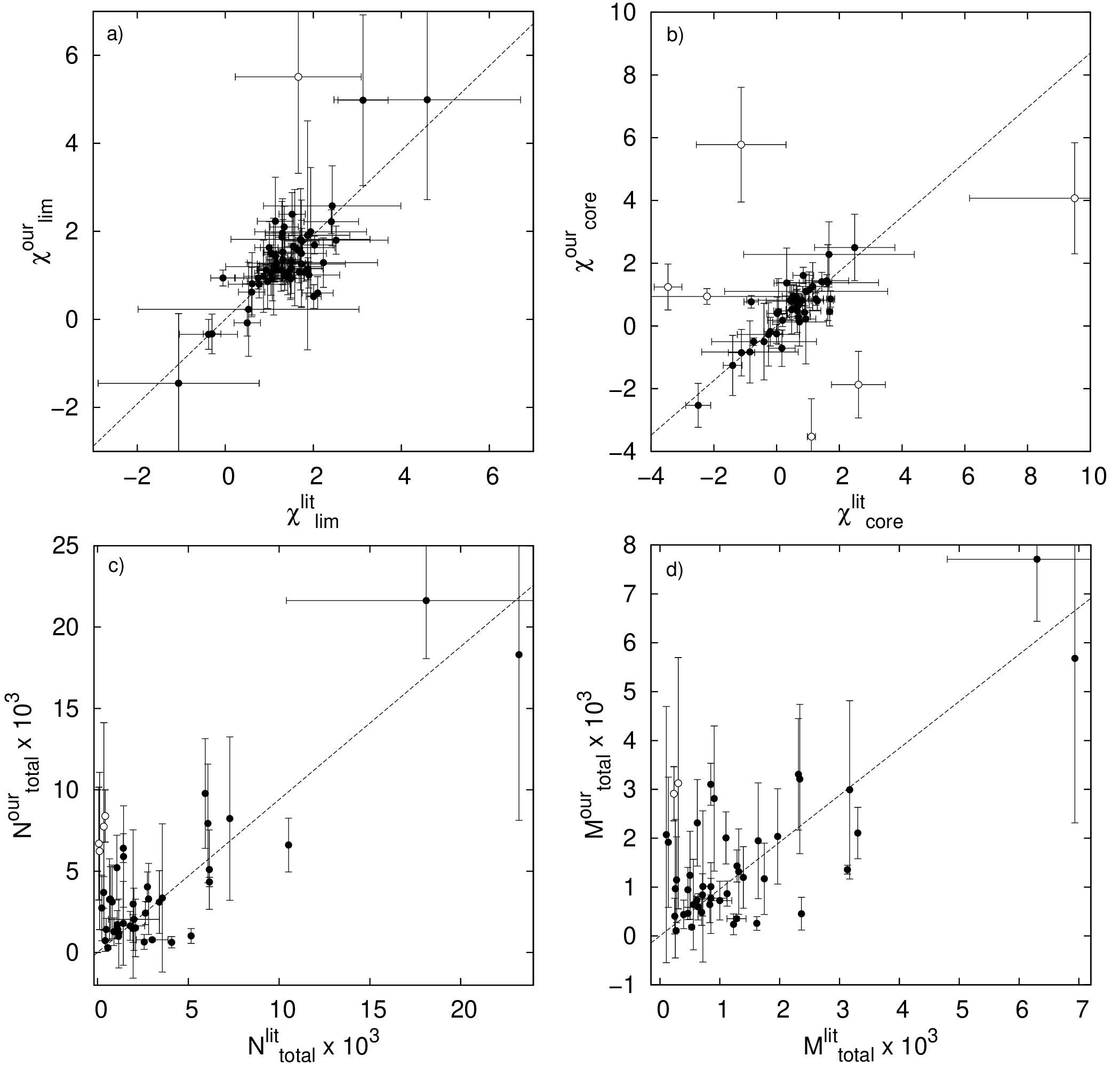}
\end{center}
\FigCap{$a)$ The MF slope for overall derived by us as a function of the literature data, with errors. The best-fitting linear relation of the form $y = ax$ is marked with a dashed line. Removed outlying points are marked as open circles. $b)$ This same as $panel\:\: a$ but for the core MF. $c)$ This same as $panel\:\: a$ but for the total number of stars. $d)$ This same as panel a but for the total mass. See Sect.~3 for details.}
\end{figure}
with the correlation coefficients of 0.83 and 0.85, respectively. In panel $c$ of Fig.~7 we present a comparison  of the total number of stars in each open clusters. For 44 objects we received the linear relation:
\begin{equation}
\centering
N^{\:our}_{\:total} = (0.94 \pm 0.09)\: N^{\:lit}_{\:total}\:
\end{equation}
with the correlation coefficient equal to 0.70. Panel $d$ of Fig.~7 shows a comparison of the total mass of each open clusters in the solar masses. For 46 objects we obtained the linear relation:
\begin{equation}
\centering
M^{\:our}_{\:total} = (0.96 \pm 0.09)\: M^{\:lit}_{\:total}\:
\end{equation}
with the correlation coefficient of 0.66. This illustrates that our analysis of the MF is reliable.

\section{Parameter uncertainties}
We discussed the derived parameter uncertainties in section 4.2 of Paper I. Additionally, analyzing the MF, we noticed that the precision of the obtained $\chi$ parameter depends on the depth of data span between the turnoff and the Completeness Limit of the 2MASS photometry (15.8 mag in the $J$ band). We analyzed open clusters where this depth was at least 2 mag, but for some of these objects our measurements exhibit relative uncertainties greater than 100\%. We rejected such cases from further consideration (open circles in all Figures). An interval of 3 mag range or higher would allow to increase the precision of a determinations. However, limiting over considerations to such clusters would lead to a significant decrease in the number of objects.

While determining the relaxation time we assumed a typical value of the velocity dispersion equal to 3 km s$^{-1}$ (Binney \& Merrifield 1998). For some clusters $\sigma_{v}$  may have a different value, for example in M67 $\sigma_{v} = 0.81$ km s$^{-1}$ according to Girard \etal (1989), in Hyades $\sigma_{v}  = 0.3$ km s$^{-1}$ (Makarov \etal 2000). But even the change of velocity dispersion by an order of magnitude, causes only a slight change of the log $\tau$.

In panels $a$ and $b$ of Fig.~8 the MF slope and the mass segregation is plotted against the distance from the Sun. As one can see, there are no selection effects in our sample. Hence our study on the MF seems to be reliable. It is worth mentioning that after comparing $\chi_{lim}$ and $\Delta \chi$ with the other parameters derived in Paper I, we did not notice any relation effects in the sample neither.
\begin{figure}[h!]
\begin{center}
\includegraphics[width=1.0\textwidth]{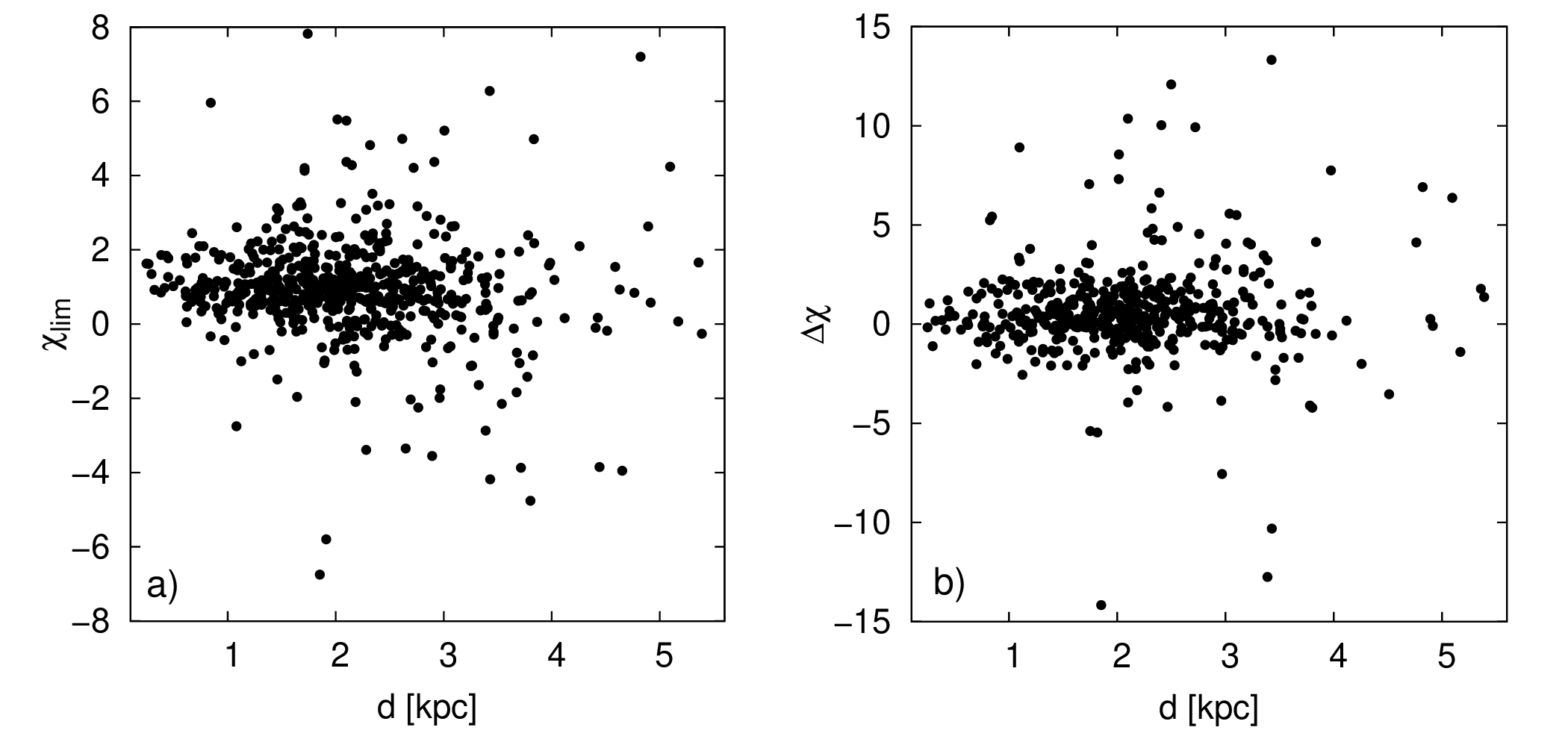}
\end{center}
\FigCap{$a)$ Relation between the MF slope and distance from the Sun. $b)$ The same but for the mass segregation parameter. See Sect.~4 for details.}
\end{figure}
\section{Conclusions}
In this paper, we analyzed the mass function, determined the initial mass function ($\chi_{0}$), the mass function slopes ($\chi$), the dynamical-evolution parameter $\tau$, the mass of a cluster and the number of member stars for 599 objects. In the case of 461 clusters these parameters were derived for cores and the coronal regions separately.

The investigation of our sample leads to the following statistical conclusions:
\begin{itemize}
\item The open clusters with small diameters do not have large masses, and vice versa. The star density in open clusters decreases with the cluster's diameter. 
\item An average open cluster reach of a medium diameter is less evolved then a small cluster with a low number of components. Thus, the dynamical-evolution parameter $\tau$ describes better evolution any status of the open clusters than just the cluster age.
\item The size of a cluster (both core and limiting) tends to decrease with the dynamic-evolution parameter $\tau$.
\item The core regions of the open clusters contains proportionally less low-mass stars then the halo areas at all cluster ages. This indicates the existence of the initial mass segregation.
\item Generally the core regions evolve faster than the halo regions so the effect of evaporation of the low-mass members is stronger there.
\end{itemize}


\Acknow{

This research is supported by UMK grant 397-F and "\emph{Stypendium ze \'srodk\'ow wspomagaj\c{a}cych, m{\l}odych naukowc\'ow oraz uczestnik\'ow studi\'ow doktoranckich na Wydziale Fizyki, Astronomii i Informatyki Stosowanej UMK}" and "Stypendia dla doktorant\'ow 2008/2009 -- ZPORR" SPS.IV-3040-UE/204/2009. This publication makes use of data products from the Two Micron All Sky Survey, which is a joint project of the University of Massachusetts and the Infrared Processing and Analysis Center/California Institute of Technology, funded by the National Aeronautics and Space Administration and the National Science Foundation.}





\begin{landscape}
\footnotetext{Full data tables available at http://www.astri.uni.torun.pl/\textasciitilde gm/OCS/2mass.html}
\renewcommand{\arraystretch}{1.0}
\renewcommand{\TableFont}{\scriptsize}
\setcounter{table}{0}
\MakeTable{lccccccccc}{13cm}{Astrophysical parameters obtained from the mass function analysis.}
{
\hline
\hline
Star cluster 	&$\chi_{lim}$&$\chi_{core}$&$\chi_{halo}$&$N_{evolved}$&$M_{turnoff}$&$N_{total}$&$M_{total}$&$N_{core}$&$M_{core}$\\
&&&&(stars)&($M_{\odot}$)&(stars)&($M_{\odot}$)&(stars)&($M_{\odot}$)\\
\hline
Berkeley 58&1.45$\pm$0.45&0.58 $\pm$ 0.40&1.35 $\pm$ 2.39&0&3.94&1769 $\pm$ 915&703 $\pm$ 364&479 $\pm$ 231&247 $\pm$ 119\\
Stock 18&0.48$\pm$0.75&--&--&2&4.6&259 $\pm$ 238&148 $\pm$ 129&--&--\\
Czernik 1&0.70$\pm$0.40&--&--&0&11.63&182 $\pm$ 93&125 $\pm$ 64&--&--\\
Berkeley 1&-0.28$\pm$6.48&0.67 $\pm$ 3.20&-2.15 $\pm$ 7.53&5&2&538 $\pm$ 21548&226 $\pm$ 8744&303 $\pm$ 1346&114 $\pm$ 498\\
King 13&1.29$\pm$0.56&1.38 $\pm$ 1.21&1.23 $\pm$ 0.70&16&2.17&13095 $\pm$ 5391&3870 $\pm$ 1815&2675 $\pm$ 2911&979 $\pm$ 1044\\
Berkeley 60&-0.18$\pm$2.28&2.75 $\pm$ 1.03&-0.78 $\pm$ 0.96&0&3.93&702 $\pm$ 9027&449 $\pm$ 5770&27 $\pm$ 60&79 $\pm$ 178\\
Mayer 1&1.21$\pm$0.79&0.81 $\pm$ 0.41&1.10 $\pm$ 0.51&0&6.21&713 $\pm$ 844&324 $\pm$ 383&279 $\pm$ 145&147 $\pm$ 76\\
King 1&2.58$\pm$0.91&1.24 $\pm$ 0.73&3.10 $\pm$ 1.10&13&1.51&6620 $\pm$ 1649&2127 $\pm$ 526&1353 $\pm$ 263&461 $\pm$ 87\\
NGC 103&1.35$\pm$1.36&0.71 $\pm$ 0.92&1.70 $\pm$ 1.53&0&4.79&2785 $\pm$ 10844&1164 $\pm$ 4531&375 $\pm$ 842&188 $\pm$ 421\\
Stock 20&0.93$\pm$0.55&0.17 $\pm$ 0.36&1.00 $\pm$ 0.71&0&5.06&867 $\pm$ 650&412 $\pm$ 309&58 $\pm$ 23&58 $\pm$ 23\\
NGC 129&1.02$\pm$0.18&0.43 $\pm$ 0.33&1.02 $\pm$ 0.17&1&5.63&2964 $\pm$ 682&1408 $\pm$ 322&136 $\pm$ 60&102 $\pm$ 45\\
Stock 21&7.82$\pm$1.28&1.60 $\pm$ 1.26&8.66 $\pm$ 1.30&4&1.69&1980 $\pm$ 972&601 $\pm$ 292&224 $\pm$ 119&77 $\pm$ 40\\
NGC 133&0.52$\pm$0.43&--&--&0&11.79&235 $\pm$ 127&218 $\pm$ 117&--&--\\
NGC 136&1.18$\pm$1.30&0.48 $\pm$ 1.82&1.84 $\pm$ 1.12&4&2.7&3526 $\pm$ 5413&1365 $\pm$ 2081&274 $\pm$ 842&119 $\pm$ 359\\
King 14&1.23$\pm$0.29&0.92 $\pm$ 0.22&1.15 $\pm$ 0.36&0&8.64&5901 $\pm$ 3117&2812 $\pm$ 1485&721 $\pm$ 251&396 $\pm$ 138\\
King 15&-1.42$\pm$2.46&-3.25 $\pm$ 2.98&-1.66 $\pm$ 2.74&3&3.41&305 $\pm$ 2741&300 $\pm$ 2641&11 $\pm$ 130&28 $\pm$ 349\\
NGC 146&1.28$\pm$0.27&0.51 $\pm$ 0.39&1.42 $\pm$ 0.30&0&6.25&5214 $\pm$ 1992&2315 $\pm$ 885&382 $\pm$ 242&268 $\pm$ 170\\
NGC 189&0.75$\pm$0.36&--&--&0&14.32&211 $\pm$ 99&150 $\pm$ 71&--&--\\
Stock 24&0.82$\pm$0.19&0.13 $\pm$ 0.43&0.87 $\pm$ 0.18&0&5.97&1440 $\pm$ 352&752 $\pm$ 184&54 $\pm$ 31&65 $\pm$ 38\\
Dias 1&0.92$\pm$0.50&1.20 $\pm$ 0.55&0.70 $\pm$ 0.43&0&9.01&726 $\pm$ 658&404 $\pm$ 366&301 $\pm$ 203&146 $\pm$ 99\\
King 16&1.14$\pm$0.29&0.64 $\pm$ 0.32&0.87 $\pm$ 0.23&0&8.6&1707 $\pm$ 875&842 $\pm$ 432&266 $\pm$ 143&175 $\pm$ 95\\
Czernik 2&0.80$\pm$0.33&0.80 $\pm$ 0.53&0.80 $\pm$ 0.27&0&15.82&978 $\pm$ 416&685 $\pm$ 292&141 $\pm$ 73&99 $\pm$ 51\\
Berkeley 4&0.94$\pm$0.25&1.78 $\pm$ 0.87&1.33 $\pm$ 0.39&0&21.87&1415 $\pm$ 688&945 $\pm$ 459&289 $\pm$ 279&120 $\pm$ 116\\
NGC 188&1.01$\pm$0.30&1.44 $\pm$ 0.20&0.74 $\pm$ 0.20&18&1.37&12200 $\pm$ 597&3902 $\pm$ 190&3084 $\pm$ 100&992 $\pm$ 32\\

\hline}
\end{landscape}
\noindent

\end{document}